\documentclass[a4paper,12pt]{article}

\newcommand{\sect}[1]{\setcounter{equation}{0}\section{#1}}

\begin{document}
\newcommand{\beq}{\begin{equation}}
\newcommand{\eeq}{\end{equation}}
\newcommand{\beqa}{\begin{eqnarray}}
\newcommand{\eeqa}{\end{eqnarray}}
\newcommand{\sr}{\sqrt}
\newcommand{\fr}{\frac}
\newcommand{\mn}{\mu \nu}
\newcommand{\G}{\Gamma}
\topmargin 0pt
\oddsidemargin 0mm

\renewcommand{\thefootnote}{\fnsymbol{footnote}}
\newcommand{\g}{\bf G_n}
\begin{titlepage}
\begin{flushright}
INJE-TP-01-10 \\
hep-th/0111136
\end{flushright}

\vspace{5mm}
\begin{center}
{\Large \bf Cosmology with the CFT-radiation matter\footnote{Talk at the 7th Italian-Korean
Symposium on Relativistic Astrophysics, Kimhae, Korea, 23-28 July 2001} }
\vspace{12mm}

{\large
 Yun Soo Myung\footnote{email address:
 ysmyung@physics.inje.ac.kr}}

\vspace{8mm}
{ Relativity Research Center and School of Computer Aided Science, Inje University,
   Kimhae 621-749, Korea}
\end{center}
\vspace{5mm}
\centerline{{\bf{Abstract}}}
\vspace{5mm}

We review  the relation between  entropy bounds to rewrite Friedmann equation
on the brane in terms of three entropy bounds:
 Bekenstein-Verlinde ($S_{BV}$); Bekenstein-Hawking
bound ($S_{BH}$); Hubble bound ($S_H$).
For a strongly coupled conformal field theory (CFT) with a dual
5-dimensional anti de Sitter Schwarzschild (AdSS$_5$) black hole, we
 can easily establish the connection between the Cardy-Verlinde formula on the CFT side and
the  entropy representation of Friedmann equation in cosmology. In this case
its cosmological evolution for  entropy  is
given by the semi-circle. However, for the matter-dominated case,
we find that the  cosmological evolution diagram takes a different form of the cycloid.
Here  we propose  two different  entropy  relations for matter-dominated case.
It turns out that the Verlinde's entropy relation so restricted that it may not  be  valid for
the matter-dominated universe.
\end{titlepage}

\newpage
\renewcommand{\thefootnote}{\arabic{footnote}}
\setcounter{footnote}{0}
\setcounter{page}{2}

\sect{Introduction}

Recently Verlinde have made two interesting things~\cite{Verl}.
The related issues appeared in~\cite{Lin,Noji,Wang,Brus,KPS,Youm}.
First he  proposed that using the  AdS/CFT correspondence~\cite{MAL}, the entropy of a
conformal field theory (CFT) in any dimension can be expressed in terms of
a generalized form of the  Cardy formula~\cite{Card}. We consider a  CFT
residing in an $(n+1)$-dimensional spacetime with the  static metric
for  the Einstein space
\beq
\label{1eq1}
 ds^2_{CFT} =-d\tau^2 +R^2d\Omega_n^2,
\eeq
where $d\Omega^2_n$ denotes  an unit $n$-dimensional sphere.
Especially for a strongly coupled CFT with
the anti de Sitter (AdS) dual, one obtains the Cardy-Verlinde
formula which states a relation between  entropy ($S$) and energy
($E$)
\beq
\label{1eq2}
S=\frac{2\pi R}{n}\sqrt{E_c(2E-E_c)}
\eeq
with the Casimir energy ($E_c$) for a finite system.
Indeed, this formula was checked to hold for various kinds of AdS-bulk spacetimes:
AdS-Schwarzschild black holes~\cite{Verl}; AdS-Kerr black holes~\cite{Klem};
 AdS-charged black holes~\cite{Cai1};  AdS-Taub-Bolt
spacetimes~\cite{Birm1,Witten2}.

The other is to   connect the above Cardy-Verlinde formula with  the $(n+1)$-dimensional
Friedmann equation based on the Friedman-Robertson-Walker (FRW) metric for the closed
universe
\begin{equation}
\label{1eq3}
ds^2_{FRW} =-d\tau^2 +{\cal R}^2(t)d\Omega_n^2.
\end{equation}
Although  two metrices  (\ref{1eq1}) and (\ref{1eq3}) have different natures,
these  are conformally equivalent.
Hence it is possible to make connection between these.
 For a radiation-dominated
closed universe, two Friedman equations are given by
\begin{eqnarray}
\label{1eq4}
&& H^2 =\frac{16 \pi G_{n+1}}{n(n-1)}\frac{E}{V} -\frac{1}{{\cal R}^2}, \\
\label{1eq5}
&& \dot{H}=-\frac{8 \pi G_{n+1}}{n-1}\left (\frac{E}{V} +p\right) +\frac{1}{{\cal R}^2},
\end{eqnarray}
where $H=\dot{{\cal R}}/{\cal R}$ is the Hubble parameter, the dot stands for the
differentiation with respect to the proper time $\tau$, $E$ is the total
energy of matter filling in the universe, $p$ denotes the pressure and
$V= {\cal R}^n \mbox{Vol}(S^n)$ is the volume of the universe. In addition, $G_{n+1}$
is the $(n+1)$-dimensional newtonian constant. Verlinde pointed out that
the Friedmann equation~(\ref{1eq4}) can be expressed in terms of  three cosmological
entropy bounds:\footnote{In Ref.~\cite{Verl}, the first bound is called
the Bekenstein bound. In fact, this bound is  slightly  different from
the original Bekenstein entropy bound proposed in~\cite{Beke,Bous}. So we call this the
Bekenstein-Verlinde bound.}

\begin{equation}
\label{1eq6}
\begin{array}{rl}
\mbox{Bekenstein-Verlinde bound}: &
S_{ BV}=\frac{2\pi}{n}E {\cal R}, \\
\mbox{Bekenstein-Hawking bound}: &
S_{ BH}=(n-1)\frac{V}{4G_{n+1}{\cal R}}, \\
\mbox{Hubble  bound}: &
S_{ H}=(n-1)\frac{HV}{4G_{n+1}}.
\end{array}
\end{equation}
Then the Friedmann equation~(\ref{1eq4}) can  be rewritten as
 the Verlinde's entropy relation
\begin{equation}
\label{1eq7}
S^2_{H}+ (S_{BV}-S_{BH})^2=S_{BV}^2.
\end{equation}
The above equation can be solved by introducing the conformal
time coordinate $\eta$ as
\begin{equation}
\label{1eq8}
S_{H}= S_{BV} \sin \eta, ~~~S_{BH}= S_{BV} (1- \cos \eta).
\end{equation}
This means that $S_{BV}$ is constant with respect to the cosmic
time $\tau$, while $S_H$ and $S_{BH}$ depend on the cosmic time.
Actually $S_{BV}$ is constant throughout the entire evolution,
because $E \sim {\cal R}^{-1}$ for a radiation-dominated universe.
We note that the Bekenstein-Verlinde bound is valid for the weakly self-gravitating
universe ($H {\cal R}\le 1$), while the Hubble bound holds for the strongly
self-gravitating universe ($H {\cal R} \ge 1$).
To decide whether a system is strongly or weakly gravitating, we
have to introduce another quantity like $S_{BH}$. When $S_{BV} \le
S_{BH}$, the system is weakly gravitating, while for $S_{BV} \ge
S_{BH}$ the self-gravity is strong. This is identified with the
holographic Bekenstein-Hawking entropy of a black hole with the
size of the universe. It grows like an  area instead of the
volume. Also the maximal entropy inside the universe is bounded
by the black holes of the size of the Hubble horizon. This is the
Hubble entropy  bound $S_H$.
It is clear from the Friedmann
equation~(\ref{1eq4}) that at the critical point of $H{\cal R}=1$, three
entropy bounds coincide exactly with each other.

Further let us propose $E_{BH}$ corresponding to the Bekenstein-Hawking
energy by using the Bekenstein-Verlinde bound such a way that $S_{BH}=(n-1)
V/4G_{n+1} {\cal R}\equiv 2\pi E_{ BH} {\cal R}/n$.
Equation~(\ref{1eq7}) then takes the form
\begin{equation}
\label{1eq9}
S_H=\frac{2\pi {\cal R}}{n}\sqrt{E_{ BH}(2E-E_{BH})}.
\end{equation}
It is very important to note here that this relation is the same form
of
the Cardy-Verlinde formula~(\ref{1eq2}) except that the roles of the entropy $(S)$ and
 Casimir energy $(E_c)$ are taken over by the Hubble entropy bound $(S_H)$ and
 Bekenstein-Hawking energy $(E_{BH})$.
This connection between the Cardy-Verlinde
formula and the Friedmann equation can be interpreted   as a consequence of the
holographic principle~\cite{Verl}.
This implies that two (Friedman  equation and Cardy-Verlinde formula)
can be derived from the same first principle.

In this direction, Savonije and Verlinde~\cite{Savo} have studied a
concrete model
by using the one-side brane
cosmology in the background of $(n+2)$-dimensional AdS-Schwarzschild
spacetime
\begin{equation}
ds^{2}_{AdSS_{n+2}}= g_{MN}dx^Mdx^N= -h(r)dt^2 +\frac{1}{h(r)}dr^2 +r^2
\left[d\chi^2 + \sin^2 \chi(d\theta^2+ \sin^2 \theta d\phi^2)
\right],
\label{BMT}
\end{equation}
where $h(r)$ is  given by
\begin{equation}
h(r)=1-\frac{m}{r^{n-1}}+ \frac
{ r^2}{\ell^2}
\end{equation}
with $\omega_{n+1} M=\frac{16 \pi G_{n+2} }{n{\rm Vol}(s^n)} M$. Here $\ell$ is
the curvature radius of AdS$_{n+2}$ space and $M$ is the ADM mass of
the black hole as measured by an observer who uses $t$ as his time
coordinate. $G_{n+2}$ is the bulk newtonian constant. In the
one-side brane world scenario, we have the relation of $G_{n+2}=\ell G_{n+1}/(n-1)$
between the bulk and boundary constants.
In the case of $m=0$, we have an exact AdS$_{n+2}$-space.  However,
$m \not=0$ generates the electric part of the Weyl tensor
$E_{00}=C_{0N0Q}n^Nn^Q \sim m/r^{n-1}$~\cite{SMS}.
This corresponds to the nonlocal effects arisen from the free
gravitational field in the bulk~\cite{Col}, transmitted through the
projection $E_{MP}$ of the bulk Weyl tensor. This nonlocal Weyl term will contributes
corrections to the Friedmann equations on the brane. Actually we
focus on the role of this term in cosmology.
It was argued that the energy $(E)$, entropy $(S)$, and temperature $(T)$ of a CFT at
high temperature can be  related to the mass $(M)$, entropy $(S_{BH})$, and Hawking
temperature $(T_H)$ of the AdS black hole. According to
the GKPW prescription of the AdS/CFT correspondence~\cite{GKPW,MAL}, the conformal
class of the boundary CFT metric is not fixed. Let us introduce
its boundary metric from the bulk one in Eq.(\ref{BMT})
\begin{equation}
ds^{2}_{BCFT}= \lim_{r \to \infty}\Big[\frac{\ell^2}{r^2}
ds^2_{AdS-S} \Big]= -d\tau^2 + \ell^2 d\Omega^2_n.
\label{CBOUN}
\end{equation}
From this we deduce the static relation between  the quantities on the
CFT-boundary
and those in the AdS bulk :
$t \to \tau=t \ell/r; T_H \to T =T_H\ell/r; M \to E =  M\ell/r.$
But we have the same entropy  $S=S_{BH}= \fr{r^n_+ {\rm Vol}(S^n)}{4 G_{n+2}},$
where $r_+$ is the event horizon of the AdS black hole.
It is well-known that the equation governing the motion of
the brane (Moving Domain Wall: MDW)
are exactly given by the $(n+1)$-dimensional Friedmann equation with
radiation matter\footnote{ There also exists the other brane cosmology:
BDL approach\cite{BDL1,BDL2}.}~\cite{CR,Krau,Ida}. In this case the  radiation-matter  which comes from
the nonlocal Weyl term can be identified  as a
strongly coupled CFT,
by making use  of  the AdS/CFT correspondence. Importantly, it turned out that the
Friedmann equation is exactly matched with the Cardy-Verlinde formula for
the CFT when the brane crosses the black hole horizon.

On the other hand authors in~\cite{Kuta,Sio} pointed out that in general,
the Cardy-Verlinde formula is not valid in weakly coupled CFTs.
Further we wish to comment that  the entropy relation of Eq.(\ref{1eq7})
is suitable only for the restricted case such as a radiation-dominated CFT.

In this article we will clarify again that a deep connection between the Cardy-Verlinde formula
and Friedmann equation  is just a
peculiar property of the dynamic brane  moving under the ``5D" anti de Sitter
Schwarzschild black hole spacetime.  Here we provide a counter example where
this connection fails for a matter-dominated universe.
Actually all moving domain
walls in the AdSS$_5$ black hole can always take a kind of
radiation-dominated matter $(\rho \sim E/V, V={\cal R}^n {\rm Vol}(S^n))$
 which arises originally
from the nonlocal term $M/{\cal R}^4$ of the
Schwarzschild black hole through the relation $ E=M \ell/{\cal R}$.
 However, if we consider the brane which moves in the ordinary bulk matter,
  this connection  is no longer satisfied. For example,
if one considers the
moving domain wall (brane) in the bulk matter with $\rho_B \not=0,P_B=0$ which does not
include any black hole,
then  one finds  equation of state for the matter-dominated universe of  $\rho_B \sim
1/a^3$ on the brane. There is also a way to obtain a radiation-dominated CFT
from the bulk spacetime in the
framework of the BDL brane cosmology~\cite{Kim,CMO}.
%%=======================section 2================================
\sect{Brane Cosmology with MDW approach}
For definiteness we choose $n=3$ (five-dimensional
AdS-Schwarzscild balck hole spacetime).
Now we introduce the radial  location of a MDW in the form of
$ r={\cal R}(\tau),t=t(\tau)$ parametrized by the proper time $\tau$
to define a cosmic embedding:
 $(t,r,\chi,\theta,\phi)\to(t(\tau),{\cal R}(\tau), \chi, \theta, \phi)$.
Then we expect that the induced metric of dynamical domain wall will be given
by the  FRW-type.
Hence $\tau$ and ${\cal R}(\tau)$ will imply the cosmic time and scale factor of
the FRW-universe, respectively.
A tangent vector (proper velocity)  of this MDW

\beq
u= \dot t \fr{\partial}{\partial t}+ \dot {\cal R} \fr{\partial}{\partial
{\cal R}},\label{TAN}
\eeq
is introduced to define this  embedding properly.
Here  overdots mean  differentiation with respect to
$\tau$. This is normalized to  satisfy
\beq
u^{M}u^{N} g_{MN}=-1.
\label{NTA}
\eeq
Given a tangent vector $u_M$, we need a normal 1-form
 directed toward to the bulk. Here we choose this as
\beq
n= \dot {\cal R} dt-  \dot t d{\cal R}, ~~~n_{M}n_{N} g^{MN}=1.
\label{NNV}
\eeq
This convention  is consistent with the Randall-Sundrum case in the limit
of $m=0$~\cite{RS}. Using  Eq.(\ref{TAN})  either with (\ref{NTA}) or
with Eq.(\ref{NNV}), we can express the proper time rate of the AdSS$_5$
time
 $\dot t$ in terms of $\dot {\cal R}$
as
\beq
\dot t=\fr{\sqrt{\dot {\cal R}^2 +h({\cal R})}}{h({\cal R})}.
\label{TAV}
\eeq
From the above, we worry about that $\dot t$ is not defined at ${\cal R}=r_+$
because $h(r_+)=0$. This also happens in the study of static black hole.
Usually one introduces  a tortoise coordinate $r^*= \int h^{-1}dr$ to resolve it.
Then Eq.(\ref{BMT}) takes a form of $ds^2_{AdSS_5}= -h(dt^2-dr^{*2})\cdots$ and
one finds the Kruscal extension. This means that $r=r_+$ is just a coordinate
singularity. We confirm this from the computation of
$R_{MNPQ}R^{MNPQ}=40/\ell^4 + 72 m^2/r^8$,
which shows that $r=0$ $(r=r_+)$ are   true (coordinate) singularities.
It is  found that there does not exits such a problem even for the dynamic
case~\cite{Myun1}.
Using Eq.(\ref{TAV}), Eq.(\ref{BMT})  leads to the 4D induced
metric for the brane
\beqa
ds^{2}_{FRW}&&=-d \tau^2 +{\cal R}(\tau)^2
\left[d\chi^2 + \sin^2 \chi (d\theta^2+ \sin^2 \theta d\phi^2)
\right]
\nonumber \\
&&\equiv h_{\mu \nu}dx^{\mu} dx^{\nu},
\label{INM}
\eeqa
where we use the Greek indices only  for the brane.
Actually the embedding of the FRW-universe into  AdSS$_5$ space is a
$2(t,r) \to 1(\tau)$-mapping. The  projection tensor is given by $h_{MN}=g_{MN}-n_Mn_N$
and its determinant is zero. Hence its inverse  $h^{MN}$ cannot be
defined. This means that the above embedding belongs to a peculiar
mapping to obtain the induced metric $h_{\mu\nu}$ in  the AdSS$_5$ black
hole spacetime $g_{MN}$ with $n_M$. Now we have to calculate the
scale factor ${\cal R}(\tau)$ from the Israel junction condition
by introducing the extrinsic curvature~\cite{ISR}.
For this case the extrinsic curvature is defined  by

\beqa
&&K_{\tau\tau}=K_{MN} u^M u^N =(h({\cal R}) \dot t)^{-1}(\ddot {\cal R} +h'({\cal R})
 /2)=\fr{\ddot {\cal R} +h'({\cal R})/2}
{\sqrt{\dot {\cal R}^2 +h({\cal R})}}, \\
&&K_{\chi\chi} = K_{\theta\theta}=K_{\phi\phi}
=- h({\cal R}) \dot t {\cal R}=-\sqrt{\dot {\cal R}^2 +h({\cal R})}~{\cal R},
\eeqa
where  prime stands for  derivative with respect to ${\cal R}$.
A localized matter on the brane implies that the extrinsic curvature jumps  across the brane.
This jump is  described  by the Israel junction condition for the
one-side brane world scenario~\cite{Myun1,Myun2}

\beq
K_{\mu \nu}=-\kappa^2_5 \left(
T_{\mu\nu}-\fr{1}{3}T^{\lambda}_{\lambda}h_{\mu\nu} \right)
\label{4DI}
\eeq
with $\kappa^2=8 \pi G_5$.  For cosmological purpose,
we may  introduce  a localized stress-energy tensor on the
brane as the 4D perfect fluid

\beq
T_{\mu \nu}=(\varrho +p)u_{\mu}u_{\nu}+p\:h_{\mu\nu}.
\label{MAT}
\eeq
Here $\varrho=\rho_m+ \sigma$ $(p=p_m-\sigma)$,
where $\rho_m $ $(p_m)$ are  the energy density (pressure)
of the localized matter and $\sigma$ is the brane tension.
Here we consider the cosmological evolution without any localized
matter on the brane. This means that there exists only the matter from the bulk
configuration  on the brane.
In this case of  $\rho_m=p_m=0$, the r.h.s. of
Eq.(\ref{4DI}) leads to  a form of the RS case as $-\fr{\sigma \kappa^2}{3}
h_{\mu\nu}$\cite{CH,Tye,Myun3}.
From Eq.(\ref{4DI}), one finds
 the space component of the junction condition

\beq
\sqrt{h({\cal R}) + \dot {\cal R}^2}=\fr{\kappa^2}{3}\sigma {\cal R}.
\label{SEE}
\eeq
For the one-side AdSS$_5$ space,
 we have the  brane tension $\sigma=3/(\kappa^2\ell)$ for the fine-tuning.
The above equation   leads to
\beq
H^2=- \fr{1}{{\cal R}^2} +\fr{m}{{\cal R}^4},
\label{HHH}
\eeq
where  $m/{\cal R}^4$
originates from the electric (Coulomb) part of the 5D Weyl tensor, $E_{00} \sim
m/r^2$~\cite{SMS,MSM}.
For $n=3$, we have  $m=\fr{16 \pi G_5 M}{3 V(S^3)},M=\fr{{\cal R}}{\ell}E,
 V={\cal R}^3 {\rm Vol}(S^3), G_5=\fr{\ell}{2} G_4$. Then one finds a CFT-radiation dominated
universe
\beq
H^2=- \fr{1}{{\cal R}^2} +\fr{8\pi
G_4}{3}\rho_{CFT},~~~\rho_{CFT}=\fr{E}{V}.
\label{CRA}
\eeq
It seems  that the equation (\ref{SEE}) is well-defined even at
${\cal R}=r_+$. Thus this leads to $H= \pm 1/\ell$ at the horizon, which
was the case mentioned first in ref.~\cite{Savo}. At this moment when the brane
crosses the horizon of the AdSS$_5$ black hole, we find that the
entropy density $s=S/V$ and the temperature of the CFT can be
expressed in terms of  the Hubble parameter $H$ and its time derivative $\dot H$ only
\beq
s= \fr{H}{2G_4},~~~ T=- \fr{\dot H}{2 \pi H},~~~ {\rm at}~ {\cal
R}=r_+.
\label{2eq13}
\eeq

Now let us discuss thermodynamics of the CFT itself.
Furthermore from the first law of thermodynamics ($TdS=dE +PdV$) and the
CFT-radiation matter ($\rho_{CFT}=M\ell/({\cal R}V), P_{CFT}=\rho_{CFT}/3$),
we derive
\beq
\fr{3}{2}(\rho_{CFT} +P_{CFT} -sT)=\fr{\gamma}{{\cal
R}^2}
\label{2eq14}
\eeq
with
\beq
~~~T=\fr{1}{ 4 \pi {\cal R}}\Big(\fr{4r_+}{\ell}+
\fr{2\ell}{r_+}\Big),~~~\gamma=\fr{3}{8 \pi G_4}\fr{r_+^2}{{\cal R}^2}.
\label{2eq15}
\eeq
Here the r.h.s of Eq.(\ref{2eq14}) represents the geometric
Casimir part of the energy density. Also from the $(3+1)$-dimensional Cardy-Verlinde formula
Eq.(\ref{1eq2}), we find
\beq
s^2=\Big(\fr{4 \pi}{3}\Big)^2\gamma \Big(\rho_{CFT}- \fr{\gamma}{{\cal
R}} \Big).
\label{2eq16}
\eeq
Two of Eqs.(\ref{2eq14}) and (\ref{2eq16}) are valid at
all times. Let us check what happens for these at the moment when the MDW crosses
the horizon. In this case we have $\gamma=\fr{3}{8 \pi G_4}$.
Using this and Eq.(\ref{2eq13}), we can recover the first Friedmann equation
(\ref{1eq4}) from Eq.(\ref{2eq16}). Also the second Friedmann
equation (\ref{1eq5}) can be derived from (\ref{2eq14}).
These imply that the Friedmann equations know about thermodynamics
of the CFT.

For a charged background, we find the same result except the
appearance of the negative energy density~\cite{BM,Myun2}. The
similar result was found for the dilatonic black hole
background~\cite{CZ}.

\sect{Entropy bound relation revisited}
In this section we focus on  the $n=3$ case. The relation for the entropy
bounds Eq.(\ref{1eq7}) corresponds to just an algebraic version  of the
radiation-dominated FRW equation (\ref{1eq4}). In other words,
this relation  reflects partly the nature of a kind of newtonian
equation :
\begin{equation}
\label{3eq1}
\dot {\cal R}^2 + V_r({\cal R})=-1
\end{equation}
with
\begin{equation}
\label{3eq2}
V_r({\cal R})= -\frac{{\cal R}_r^2}{{\cal R}^2},~~ {\cal R}_r^2=
\frac{4 \pi G_4 E}{3{\rm Vol}(S^3)}.
\end{equation}
The solution of this differential equation can be solved parametrically in
terms of an arc parameter $\eta$~\cite{MTW},
\begin{equation}
\label{3eq3}
{\cal R}= {\cal R}_r \sin \eta, ~~~\tau= {\cal R}_r (1- \cos \eta).
\end{equation}
The range of $\eta$ from start of expansion to end of
recontraction  is just $\pi$ and the curve relating  radius ${\cal R}$ to time
$\tau$
is a semicircle. From Eq.(\ref{3eq3}), we find a  relation
representing the diagram for  cosmic evolution
\begin{equation}
\label{3eq4}
{\cal R}^2+ ({\cal R}_r- \tau)^2={\cal R}_r^2
\end{equation}
which is the same form as in the entropy relation Eq.(\ref{1eq7}).
Of course this is valid  for the strongly self-gravitating universe with
$H{\cal R}>1$. In this case we expect a naive correspondence between $({\cal R},\tau,{\cal R}_r)$
and $(S_H,S_{BH},S_{BV})$:
\begin{equation}
\label{3eq5}
{\cal R} \leftrightarrow S_H, ~~ \tau \leftrightarrow S_{BH},
 ~~ {\cal R}_r \leftrightarrow S_{BV}.
\end{equation}
Also, we observe that for a radiation-dominated universe,
 the solution  (\ref{3eq3}) to the
differential equation  takes exactly the same form as in  Eq.(\ref{1eq8}) to the
algebraic equation for the entropy bounds. This property can be
regarded as an important factor for establishing the connection between the Cardy-Verlinde
formula  and the FRW-equation in the radiation-dominated CFT universe.
In the next we introduce a counter example where this connection
is no longer satisfied.

%%====================section 4=================================
\sect{Matter-dominated universe}

In this section we consider the moving domain wall in the the AdS$_5$ spacetime
with the negative cosmological constant $\Lambda$ including the other bulk
matter $(\tilde M)$. We assume that the AdS space does not include any object like black hole.
We introduce the matter-dominated Friedmann equation
\begin{equation}
\label{4eq1}
\dot {\cal R}^2 + V_m({\cal R})=-1
\end{equation}
with
\begin{equation}
\label{4eq2}
V_m({\cal R})=- \frac{{\cal R}_m}{{\cal R}},~~ {\cal R}_m=
 \frac{8 \pi G_4 \tilde M}{3{\rm Vol}(S^3)}.
\end{equation}
This situation may be figured out from the brane world
scenario~\cite{Kim,CMO}. This can  be derived  from the newtonian
cosmology~\cite{Dau}.
For example, we consider a spherical ball $(S^3)$ on the brane
with  matter $m$
in it. Of course this arises from a kind of the bulk matter $\rho_B \sim \tilde M,~p_B=0$.
In this case the potential of the ball on its surface  is
given by $\phi_{ball}(r={\cal R}(\tau))= -G_4 m/{\cal R}$. Let us introduce a
point-like probe with unit mass on the ball. The energy
conservation condition for the bound-motion with negative total energy $-k/2$ is
given by
\begin{equation}
\label{4eq3}
\frac{\dot {\cal R}^2}{2} + \phi_{ball}=-\frac{k}{2}
\end{equation}
which can be led to the matter-dominated FRW-equation Eq.(\ref{4eq1})
if
one chooses $m=4 \pi \tilde M/3 {\rm Vol}(S^3)$ and $k=1$.
However, one handicap of the newtonian
cosmology is that this choice of $m,k$  is unclear.

The solution to the equation (\ref{4eq1}) can be expressed parametrically in
terms of an arc parameter $\eta$~\cite{MTW},
\begin{equation}
\label{4eq4}
{\cal R}= \frac{{\cal R}_m}{2}(1- \cos\eta), ~~~\tau= \frac{{\cal R}_m}{2} (\eta- \sin \eta).
\end{equation}
Here the range of $\eta$ from start of expansion to end of
recontraction  is  $2\pi$ and the curve relating  radius ${\cal R}$ to time
$\tau$
is not a semicircle. These are  different points when comparing
with the radiation-dominated case. From Eq.(\ref{4eq4}), we find a
 relation between ${\cal R}$ and $\tau$ expressed as cycloid, compared with a semicircle
for the radiation-dominated universe
Eq.(\ref{3eq4})
\begin{equation}
\label{4eq5}
({\cal R}_m-2 {\cal R})^2+ ({\cal R}_m \eta -2 \tau)^2={\cal R}_m^2.
\end{equation}
This is valid  for the weakly self-gravitating universe with
$H{\cal R}<1$. The newtonian cosmology also belongs to this category.
Assuming a naive correspondence  between $({\cal R},\tau,{\cal R}_m)$
and $(S_H,S_{BH},S_{BV})$ as in the radiation-dominated case,
then one finds one entropy relation from Eq.(\ref{4eq5}) :
\begin{equation}
\label{4eq6}
(S_{BV}-2 S_H)^2+ (S_{BV} \eta - 2 S_{BH})^2=S_{BV}^2,
\end{equation}
where $S_{BV}$ is constant.

On the other hand, if one follows closely the definition of each
entropy bound,  we may propose the  new entropy  bounds : Bekenstein-Verlinde,
Bekenstein-Hawking, and Hubble bounds which may be useful for describing
 the matter-dominated universe
\begin{equation}
\label{4eq7}
\begin{array}{rl}
\mbox{Bekenstein-Verlinde bound}: &
\tilde S_{ BV}=\frac{2\pi}{n}\tilde M {\cal R}, \\
\mbox{Bekenstein-Hawking bound}: &
\tilde S_{ BH}=(n-1)\frac{V}{4G_n{\cal R}}, \\
\mbox{Hubble  bound}: &
\tilde S_{ H}=(n-1)\frac{HV}{4G_n}.
\end{array}
\end{equation}
These are  the same forms as in Eq.(\ref{1eq6}) except for the definition
of the Bekenstein-Verlinde bound. In the radiation-dominated CFT case,
$S_{BV}$ is constant because of $E{\cal R}=M\ell$(=constant), whereas
for the matter-dominated universe, $\tilde S_{BV}$ scales as ${\cal R}$ as the
universe evolves. This means that $\tilde S_{BV}$  is not constant here. But we
suggest  a familiar relation between new entropy bounds
\begin{equation}
\label{4eq8}
\tilde S^2_H+ (\tilde S_{BV}-\tilde S_{BH})^2=\tilde S_{BV}^2.
\end{equation}
As it stands, this  is different from Eq.(\ref{1eq7}).
Eq.(\ref{4eq8})
  is not obviously an equation for the semicircle because $\tilde S_{BV}$
is not constant.
\sect{Conclusions}
At this time we do not know exactly which one among Eqs.(\ref{4eq6}) and (\ref{4eq8})
is appropriate  for describing
the entropy relation for the matter-dominated universe. Also we note that the Verlinde's
entropy relation Eq.(\ref{1eq7}) which derives  from  the first Friedmann
equation  Eq.(\ref{1eq4}) for the radiation-dominated case
 is not suitable for the matter-dominated
case. Finally we wish to comment that a close relationship between the
Cardy-Verlinde formula and Friedmann equation is realized only for  a
special circumstance such as a radiation-dominated CFT within the AdS/CFT
correspondence through the  brane cosmology.

\section*{Acknowledgments}
We thank to R.G. Cai and  H. W. Lee  for helpful discussions.
This  was supported
in part by the Brain Korea 21 Program, Ministry of Education, Project No.
D-0025 and KOSEF, Project No. 2000-1-11200-001-3.

%\newpage
%%=======================references===========================

\end{document}